\documentstyle{article}
\title{A relationship between scalar Green functions on hyperbolic and Euclidean Rindler spaces }
\author{ Z. Haba\\
Institute of Theoretical Physics, University of Wroclaw,\\ 50-204
Wroclaw, Plac Maxa Borna 9, Poland,\\email:zhab@ift.uni.wroc.pl}
\begin{document}\maketitle
\date{}
\begin{abstract}
 We derive a formula connecting in any dimension the
Green function on the D+1 dimensional Euclidean Rindler space and
the one for a  minimally coupled scalar field with a mass $m$ in
the D dimensional hyperbolic space.
 The relation takes a simple form in the
momentum space
 where the Green functions are equal
at the momenta $(p_{0},{\bf p})$ for Rindler and $(m,\hat{{\bf
p}})$ for hyperbolic space with a simple additive relation between
the squares of the mass and the momenta. The formula has
applications to finite temperature Green functions, Green
functions on the cone and on the (compactified) Milne space-time .
Analytic continuations and interacting quantum fields  are briefly
discussed.\end{abstract}
 \section{Introduction}The Green functions
define quantum field theory in the Minkowski space as well as in
the curved space. In the Minkowski space the relation is one to
one if we impose the restrictions of locality and Poincare
invariance. In the curved space  the Green function is not unique.
This is a consequence of the  non-uniqeness of the physical vacuum
\cite{fulling}\cite{milne}. There is less ambiguity for the Green
functions on the Riemannian manifolds (instead of the physical
pseudo-Riemannian ones). However,  in the Euclidean approach
\cite{wald}\cite{dimock}\cite{jaffe} we can construct only a
subset of quantum field theories admissible on the
pseudo-Riemanian manifolds. The Euclidean version on a manifold
can have more than one analytic continuation. The hyperbolic space
can be continued analytically either to de Sitter space or to anti
de Sitter space. In the case of the de Sitter and anti de Sitter
spaces the relation between the Riemannian  and pseudoRiemannian
approaches is well understood (the Euclidean approach is
distinguishing the "Euclidean vacuum" also known as the
"Bunch-Davis vacuum" \cite{bunch}).

In this paper we discuss Green functions on the Euclidean version
of the Rindler space and on the hyperbolic space . The Euclidean
Rindler space continues analytically either to the conventional
Rindler space of an accelerated observer \cite{rindler} or to the
Milne space \cite{milne} which plays an important role in the
"ekpyrotic" scenario \cite{ekpy}.
 De Sitter and anti de Sitter
spaces can be considered as asymptotic solutions (near the
singularity) of gravity and string models (black branes
\cite{horowitz}, see also \cite{maldacena}) . The Rindler space
describes near the horizon geometry of space-time .
 We find that after a Fourier transformation
in time the massless Green function in the Rindler space is equal
to the
 Green function in the hyperbolic space for a quantum field with a mass $m$
  (related to
 $p_{0}$). The relation can be
extended to the  Rindler Green function with mass $m_{R}$ but in
such a case we have to take the Fourier transform in spatial
coordinates as well. Then, the Fourier transforms of both Green
functions are equal for  momenta $(p_{0},{\bf p})$ (Rindler) and
$(m,\hat{{\bf p}})$ (hyperbolic).

The Green functions on the Rindler space have been discussed by
many authors
\cite{fulling}\cite{boulware}\cite{vandam}\cite{moretti1}\cite{bunch}.
A requirement of the (twisted) periodic conditions relates Green
functions on the conical manifold
\cite{dowker1}\cite{dowker3}\cite{dowker2}( a solution for the
cosmic string \cite{vilenkin} \cite{sereb}) to the Rindler Green
functions and Rindler quantum fields at finite temperature (then
there is no twist). These Green functions have been studied by
many authors (mainly in four dimensions) by means of an
eigenfunction expansion \cite{japan}\cite{linet1}\cite{linet2}
\cite{moreira}. It seems however that the simple relations derived
in this paper have not been known before.
 The Green functions and quantum fields on de Sitter and anti de
Sitter spaces have been discussed in many physics papers
\cite{fronsdal}
\cite{ford}\cite{schom}\cite{mottola}\cite{allen}\cite{allenfolacci}
\cite{kirsten}\cite{tsamis}\cite{bros} \cite{maldacena} as well as
in mathematics \cite{rycher}\cite{gaveau}\cite{anker}.

\section{Green functions}
We are interested in metrics with a bifurcate Killing horizon in
$D+1$ dimensional pseudoRiemannian manifold ${\cal N}$. This
notion is defined  (see ref.\cite{bif} for details) by a Killing
vector vanishing on a $D-1$ dimensional surface. The bifurcate
Killing horizon  locally divides the space-time  into four wedges
as the boost Killing field does it in the case of the Minkowski
space \cite{rindler}\cite{bif}. We take as our starting point the
Euclidean version of the Rindler approximation of the metric on
${\cal N}$
\begin{equation}
\begin{array}{l}ds^{2}\equiv
g_{AB}dx^{A}dx^{B}=y^{2}(dx^{0})^{2}+dy^{2}+d{\bf x}^{2}
\end{array}\end{equation}
 An analytic continuation $x^{0}\rightarrow
ix^{0}$ transforms the metric (1) back into the pseudo Riemannian
Rindler metric. The analytic continuation of both $
x^{0}\rightarrow ix^{0}$ and $y\rightarrow it$ transforms the
metric (1) into the Milne metric \cite{milne}.

 Let
\begin{equation}
\triangle_{N}=\frac{1}{\sqrt{g}}\partial_{A}g^{AB}\sqrt{g}\partial_{B}
\end{equation}
be the Laplace-Beltrami operator on ${\cal N}$ ( here
$g=\det(g_{AB}$)).

We are interested in the calculation of the Green functions for a
minimal coupling of the scalar field\begin{equation}
(-\triangle_{N} +m^{2}){\cal G}_{N}^{m}=\frac{1}{\sqrt{g}}\delta
\end{equation}

 A solution of eq.(3) can be expressed by the fundamental solution
of the diffusion equation\begin{equation}
\partial_{\tau}P=\frac{1}{2}\triangle_{N}P
\end{equation}
with the initial condition
$P_{0}(x,x^{\prime})=g^{-\frac{1}{2}}\delta(x-x^{\prime})$. Then

\begin{equation}
{\cal
G}_{N}^{m}=\frac{1}{2}\int_{0}^{\infty}d\tau\exp(-\frac{1}{2}m^{2}\tau)
P_{\tau}
\end{equation}
In order to prove eq.(5) we multiply eq.(4) by
$\exp(-\frac{1}{2}m^{2}\tau)$ and integrate both sides over $\tau$
applying the initial condition for $P_{\tau}$.

In the metric (1) we have
 \begin{equation}
 \triangle_{N}=y^{-2}\partial_{0}^{2}+y^{-1}\partial_{y}y\partial_{y}+\triangle
 \end{equation}
 where $\triangle$ is the Laplacian on $R^{n}$ where $n=D-1$.
After  an exponential change of coordinates
\begin{equation}
y=\exp u
\end{equation}
eq.(3) reads ( we denote the mass of the Rindler field by $m_{R}$)
\begin{equation}\begin{array}{l}
\Big(-\partial_{0}^{2}-\partial_{u}^{2}+\exp(2u)(-\triangle+m_{R}^{2})\Big){\cal
G}^{m}_{R}
=\delta(x_{0}-x_{0}^{\prime})\delta(u-u^{\prime})\delta({\bf
x}-{\bf x}^{\prime})\end{array}
\end{equation}
The metric (1) by means of the conformal transformation is related
to the metric $(dx^{0})^{2}+ds_{H}^{2}$ on $R\times {\cal H}_{D}$
where
 \begin{equation}
 ds_{H}^{2}=y^{-2}(dy^{2}+d{\bf x}^{2})
 \end{equation}
 (with ${\bf x}\in R^{D-1}$) is the Riemannian metric (the Poincare metric) on the hyperbolic space
 ${\cal H}_{D}=SO(D,1)/SO(D)$ .

 The Laplace-Beltrami operator (2) for the hyperbolic space reads
\begin{equation}
\triangle_{H}=y^{2}(\frac{\partial^{2}}{\partial
y^{2}}+\frac{\partial^{2}}{\partial
x_{1}^{2}}+....+\frac{\partial^{2}}{\partial x_{n}^{2}})
-(n-1)y\frac{\partial}{\partial y}\end{equation}
Let\begin{equation} {\cal
G}_{H}^{m}(X,X^{\prime})=y^{\frac{n}{2}}y^{\prime\frac{n}{2}}\hat{{\cal
G}}_{H}^{m}(X,X^{\prime})
\end{equation} here $X=(y,{\bf x})$.Then, $\hat{{\cal
G}}^{m}_{H}$ satisfies the equation
\begin{equation}
(-y\partial_{y}y\partial_{y}-y^{2}\triangle+\frac{n^{2}}{4}+m^{2})\hat{{\cal
G}}^{m}_{H} =y\delta(X-X^{\prime})\end{equation} In the
coordinates (7)  eq.(12) reads
\begin{equation}
(-\partial_{u}^{2}-\exp(2u)\triangle+m^{2}+\frac{n^{2}}{4})\hat{{\cal
G}}^{m}_{H}=\delta(u-u^{\prime})\delta({\bf x}-{\bf x}^{\prime})
\end{equation}
We consider also the  heat kernel for the hyperbolic space, i.e.,
the fundamental solution of the diffusion equation
\begin{equation}
\partial_{\tau}P_{\tau}^{H}=\frac{1}{2}\triangle_{H} P_{\tau}^{H}
\end{equation}
with the initial condition
$P_{0}(X,X^{\prime})=\frac{1}{\sqrt{g}}\delta(X-X^{\prime})$.

If we have the fundamental solution (14) then we can solve the
equation for the Green function
\begin{equation}
(-\triangle_{H}+ m^{2}){\cal G}_{H}^{m}=\frac{1}{\sqrt{g}}\delta
\end{equation}

 Let us define
\begin{equation}
P_{\tau}^{H}(X,X^{\prime})=y^{\frac{n}{2}}y^{\prime\frac{n}{2}}\hat{P}_{\tau}(X,X^{\prime})
\end{equation}
Then, $\hat{P}$ satisfies \begin{equation}
-\partial_{\tau}\hat{P}_{\tau}=\frac{1}{2}(-\partial_{u}^{2}-\exp(2u)\triangle+m^{2}+\frac{n^{2}}{4})\hat{P}_{\tau}
\end{equation}with the initial
condition
\begin{displaymath}
\hat{P}_{0}(X,X^{\prime})=y\delta(X-X^{\prime})
\end{displaymath}
Therefore $\hat{{\cal G}}$ defined in eq.(11) is expressed by
$\hat{P}$
\begin{equation} \hat{ {\cal G}}_{H}^{m} =\int_{0}^{\infty}d\tau
\exp(-\frac{1}{2}m^{2}\tau)\hat{P}_{\tau}
\end{equation}
We are prepared now to show a correspondence between the Green
function for the Rindler space and the Green function on the
hyperbolic space. The correspondence could be guessed on the basis
of the conformal equivalence mentioned above eq.(9)( see a
discussion of conformal invariance in \cite{candelas}
\cite{kerner}\cite{russo}). In order to prove the relationship
between the Green functions let us consider the Fourier transform
\begin{equation}\begin{array}{l}
{\cal G}_{R}^{m}(x_{0},u,{\bf x};x_{0}^{\prime},u^{\prime},{\bf
x}^{\prime}) =\frac{1}{2\pi}\int
dp_{0}\exp(ip_{0}(x_{0}-x_{0}^{\prime}))\tilde{{\cal
G}}_{R}^{m}(p_{0},u,u^{\prime};\vert {\bf x}-{\bf
x}^{\prime}\vert)\end{array}
\end{equation}
It follows from eqs.(8) and (13) that at $m_{R}=0$
\begin{equation}
\tilde{{\cal G}}_{R}^{0}(p_{0},u,u^{\prime};\vert {\bf x}-{\bf
x}^{\prime}\vert)=\hat{{\cal G}}_{H}^{m}(u,u^{\prime};\vert {\bf
x}-{\bf x}^{\prime}\vert)
\end{equation}
if
\begin{equation}
p_{0}^{2}=m^{2}+\frac{n^{2}}{4}
\end{equation}
(in this way the mass in the hyperbolic space acquires the meaning
of a momentum in the $D+1$ dimension). Applying eqs.(8),(13) and
(18) we obtain
\begin{equation}\begin{array}{l} \tilde{{\cal G}}_{R}^{0}(p_{0},u,u^{\prime};\vert
{\bf x}-{\bf x}^{\prime}\vert)
=\int_{0}^{\infty}d\tau\exp(-\frac{\tau}{2}p_{0}^{2}+\frac{\tau}{8}n^{2})\hat{P}_{\tau}
(y,{\bf x};y^{\prime},{\bf x}^{\prime})\end{array}
\end{equation}
or
\begin{equation}\begin{array}{l}
{\cal G}_{R}^{0}(x_{0},u,{\bf x};x_{0}^{\prime},u^{\prime},{\bf
x}^{\prime}) =\int_{0}^{\infty}d\tau(2\pi\tau)^{-\frac{1}{2}}
\exp(-\frac{1}{2\tau}(x_{0}-x_{0}^{\prime})^{2}+\frac{\tau}{8}n^{2})\hat{P}_{\tau}
(y,{\bf x};y^{\prime},{\bf x}^{\prime})\end{array}
\end{equation}

 The relation can be extended to $m_{R}\neq 0$. Let us denote the
 Fourier transform of ${\cal G}({\bf x})$ by $\tilde{{\cal G}}({\bf p})$. The Green function
 $\tilde{{\cal G}}({\bf p})$ depends only on
 $\vert{\bf p}\vert$. Let us denote the Fourier transform of the
 massive Rindler Green function $\tilde{{\cal
G}}_{R}^{m_{R}}(p_{0},u,u^{\prime};\vert {\bf x}-{\bf
x}^{\prime}\vert)$ in the ${\bf x}-{\bf x}^{\prime}$ variable  by
$\tilde{{\cal G}}_{R}^{m_{R}}(p_{0},u,u^{\prime};\vert {\bf
p}\vert)$  . Then,
\begin{equation}
\tilde{{\cal G}}_{R}^{m_{R}}(p_{0},u,u^{\prime};\vert {\bf
p}\vert)= \tilde{\hat{{\cal G}}}_{H}^{m}(u,u^{\prime};\vert
\hat{{\bf p}}\vert)= {\cal G}_{Q}(\vert \hat{ {\bf
p}}\vert;u,u^{\prime})
\end{equation}
if in addition to eq.(21)(expressing $m$ by $p_{0}$) the following
relation is satisfied
\begin{equation}
m_{R}^{2}+{\bf p}^{2}\equiv\omega({\bf p})^{2}=\hat{{\bf p}}^{2}
\end{equation}
(the momentum in the hyperbolic space acquires the meaning of the
 energy in the Rindler space). On the rhs of eq.(24) ${\cal G}_{Q}(\vert \hat{ {\bf
p}}\vert)$ is the integral kernel ${\cal A}_{Q}^{-1}$ of the
quantum mechanical Hamiltonian ${\cal A}_{Q}$ (with an exponential
potential)\begin{equation} {\cal
A}_{Q}=-\partial_{u}^{2}+\hat{{\bf
p}}^{2}\exp(2u)+m^{2}+\frac{n^{2}}{4}
\end{equation}
The formula (24) can be rewritten in the configuration space. For
this purpose we express $\hat{{\bf p}}$ on the rhs of eq.(24) by
$\omega({\bf p})$ from eq.(25). Then, the Fourier transform of
eq.(24) can be expressed
 by means of a kernel $K_{m_{R}}$ relating the
massive Green function to the massless one
\begin{equation}\begin{array}{l}
{\cal G}_{R}^{m_{R}}(x_{0},y,{\bf
x};x_{0}^{\prime},y^{\prime},{\bf x}^{\prime}) =\int d\hat{{\bf
x}}K_{m_{R}}({\bf x}-{\bf x}^{\prime},\hat{{\bf x}}){\cal
G}_{R}^{0}(x_{0},y,\hat{{\bf x}};x_{0}^{\prime},y^{\prime},{\bf
0})
\end{array}\end{equation}
where
\begin{equation}\begin{array}{l}
K_{m_{R}}({\bf x}-{\bf x}^{\prime},\hat{{\bf
x}})=(A(n-1))^{-1}(2\pi)^{-n} \cr\int d{\bf p}d\hat{{\bf
p}}\exp(-i\hat{{\bf p}}\hat{{\bf x}}+i{\bf p}({\bf x}-{\bf
x}^{\prime}))\delta(\vert\hat{{\bf p}}\vert- \omega({\bf
p}))\omega({\bf p})^{-n+1}\end{array}
\end{equation}and $A (n-1)$ is the area of the $n-1$-dimensional
sphere of radius 1. In order to prove eq.(27) we take the Fourier
transform of eq.(27) in ${\bf x}$. Then, the remaining integral
over $\hat{{\bf p}}$ can be performed in spherical coordinates
leading to the formula (24).
\section{Integral representation}
The heat kernel $P_{\tau}$ (14) on the hyperbolic space has been
calculated by many authors ( see \cite{campo} for a review; we
have done  the calculations by means of a probabilistic method
\cite{haba} and our results agree with those of
ref.\cite{imperial}). It is a function of the Riemannian distance
$\sigma$\begin{displaymath}\cosh \sigma\equiv
z=1+(2yy^{\prime})^{-1}(({\bf x}-{\bf
x}^{\prime})^{2}+(y-y^{\prime})^{2})\end{displaymath}
 We have for
odd dimensions $D=n+1=2k+3$ ($k=0,1,..$) (where  $P_{\tau}(y,{\bf
x};y^{\prime},{\bf x}^{\prime})\equiv
p_{\tau}(\sigma)$)\begin{equation}\begin{array}{l}
p_{\tau}^{(k+1)}(\sigma)
=(-2\pi)^{-k}\exp(-\frac{n^{2}}{8}\tau+\frac{1}{2}\tau)
\Big((\sinh\sigma)^{-1}\frac{d}{d\sigma}\Big)^{k}p_{\tau}^{(1)}(\sigma)
\end{array}\end{equation} with \begin{equation}
p_{\tau}^{(1)}(\sigma)=(2\pi
\tau)^{-\frac{3}{2}}\sigma(\sinh\sigma)^{-1}\exp(-\frac{\tau}{2}-\frac{\sigma^{2}}{2\tau})
\end{equation}
 here $(\sinh
\sigma)^{-1}\frac{d}{d\sigma}=\frac{d}{dz}$.

In even dimensions $D=n+1=2k+2$
\begin{equation}
\begin{array}{l}
p_{\tau}^{(k)}(\sigma)=\exp(-\frac{n^{2}\tau}{8}+\frac{\tau}{8})(-2\pi)^{-k}
 \Big((\sinh \sigma)^{-1}\frac{d}{d\sigma}\Big)^{k}
p_{\tau}^{(0)}(\sigma)
\end{array}
\end{equation} where \cite{mckean}

\begin{equation}\begin{array}{l}p_{\tau}^{(0)}(\sigma)=\exp(-\frac{\tau}{8})\sqrt{2}(2\pi
\tau)^{-\frac{3}{2}}\int_{\sigma}^{\infty}(\cosh r -\cosh
\sigma)^{-\frac{1}{2}}r\exp(-\frac{r^{2}}{2\tau})dr
\end{array}\end{equation}Then, the Green functions for the hyperbolic
space read ($n=2k+2,k=0,1,...$)\begin{equation}\begin{array}{l}
{\cal G}_{H}^{m}(y,{\bf x};y^{\prime},{\bf
x}^{\prime})_{2k+2}=(-2\pi)^{-k} \Big((\sinh
\sigma)^{-1}\frac{d}{d\sigma}\Big)^{k}(\sinh
\sigma)^{-1}\exp(-\nu\sigma) \end{array}\end{equation} where
\begin{equation} \nu=\sqrt{\frac{n^{2}}{4}+m^{2}}\end{equation}
and for an odd $n$
  \begin{equation} \begin{array}{l}{\cal
G}_{H}^{m}(y,{\bf x};y^{\prime},{\bf x}^{\prime})_{2k+1}\cr
 =2\sqrt{2}(2\pi)^{-\frac{3}{2}}
(-2\pi)^{-k} \Big((\sinh \sigma)^{-1}\frac{d}{d\sigma}\Big)^{k}
\int_{\sigma}^{\infty}(\cosh r -\cosh
\sigma)^{-\frac{1}{2}}\exp(-\nu r)dr\cr =2(2\pi)^{-\frac{3}{2}}
(-2\pi)^{-k} \Big((\sinh
\sigma)^{-1}\frac{d}{d\sigma}\Big)^{k}Q_{\nu-\frac{1}{2}}(\cosh
\sigma)
\end{array}
\end{equation}
where the Legendre function $Q$ has the integral
representation\cite{grad}
\begin{displaymath}
 Q_{\alpha}(z)=\int_{\sigma}^{\infty}(2\cosh r -2z)^{-\frac{1}{2}}\exp(-\frac{(2\alpha+1)r}{2})dr
\end{displaymath}

The Fourier transform in $x_{0}$ of the massless Rindler Green
function  is equal to the hyperbolic Green function with (see
eqs.(20)-(21) and (34))
\begin{displaymath}
\nu=\vert p_{0}\vert
\end{displaymath}
in eqs.(33) and (35). Explicitly, for an even case $n=2k+2 $ ($
D=n+2$) we have the simple formula

\begin{equation}\begin{array}{l}
\tilde{{\cal G}}_{R}^{0}(p_{0},y,{\bf x};y^{\prime},{\bf
x}^{\prime})_{2k+2} =(-2\pi)^{-k}y^{-k-1}y^{\prime-k-1}\cr
\Big((\sinh \sigma)^{-1}\frac{d}{d\sigma}\Big)^{k}(\sinh
\sigma)^{-1}\exp(-\vert p_{0}\vert\sigma)\end{array}
\end{equation}

The massless  Green function in $D+1=2k+4$ dimensional Rindler
space can be obtained either from eq.(36) by means of the Fourier
transform in $p_{0}$ or from eq.(23) by a calculation of the
$\tau$-integral
\begin{equation}\begin{array}{l}
{\cal G}^{0}_{R}(x_{0}-x_{0}^{\prime},y,y^{\prime};\sigma)_{2k+2}
\cr =(-2\pi)^{-k-2}y^{-k-1}y^{\prime -k-1}\Big((\sinh
\sigma)^{-1}\frac{d}{d\sigma}\Big)^{k}\sigma
(\sinh\sigma)^{-1}\Big(\sigma^{2}+(x_{0}-x_{0}^{\prime})^{2}\Big)^{-1}
\end{array}
\end{equation} ( for $k=0$ the formula has been derived in
\cite{vandam}\cite{dowker1}\cite{dowker3} and \cite{bunch2})

 In even
dimensions $D=2k+2$ the formula for the Rindler Green function is
more complicated . From eqs.(23) and (32) we obtain
\begin{equation}\begin{array}{l} {\cal G}^{0}_{R}(x_{0}-x_{0}^{\prime},y,y^{\prime};\sigma)_{2k+1} =
-(-2\pi)^{-k-1}y^{-k-\frac{1}{2}}y^{\prime
-k-\frac{1}{2}}\cr\Big((\sinh
\sigma)^{-1}\frac{d}{d\sigma}\Big)^{k}
\int_{\sigma}^{\infty}(\cosh r -\cosh
\sigma)^{-\frac{1}{2}}r(r^{2}+(x_{0}-x_{0}^{\prime})^{2})^{-1}dr
\end{array}\end{equation}

We can extend the formulas
 to quantum field theory at finite
temperature and to a construction of Green functions (with a zero
twist) on the conical manifolds \cite{dowker1}\cite{moreira}. The
Euclidean Green functions at finite temperature  are constructed
\cite{temp} from ${\cal G}$ by an imposition of the periodicity
condition in time by means of the method of images. Applying the
formula\begin{displaymath}\begin{array}{l}
\sum_{n}\Big(\sigma^{2}+(x_{0}-x_{0}^{\prime}+ n\beta)^{2}
\Big)^{-1}\cr
 =\pi(2\beta\sigma)^{-1}\Big(\coth(\frac{\pi}{\beta}(\sigma
+i(x_{0}-x_{0}^{\prime})) +\coth(\frac{\pi}{\beta}(\sigma
-i(x_{0}-x_{0}^{\prime}))\Big)\end{array} \end{displaymath} we
obtain (in four dimensions the formula has been derived by
\cite{dowker1}\cite{dowker3}\cite{moretti1}\cite{moretti2})\begin{equation}
\begin{array}{l}{\cal
G}^{0}_{2k+2}(x_{0}-x_{0}^{\prime},y,y^{\prime};\sigma)_{\beta}
=(-2\pi)^{-k-2}y^{-k-1}y^{\prime -k-1}\cr\Big((\sinh
\sigma)^{-1}\frac{d}{d\sigma}\Big)^{k}
\sinh(\frac{2\pi}{\beta}\sigma)\cr\pi(2\beta\sinh \sigma
)^{-1}\Big(\cosh(\frac{2\pi}{\beta}\sigma)
-\cos(\frac{2\pi}{\beta}(x_{0}-x_{0}^{\prime}))\Big)^{-1}

\end{array}\end{equation} and
 \begin{equation}\begin{array}{l} {\cal G}^{0}_{2k+1}(x_{0}-x_{0}^{\prime},y,y^{\prime};\sigma)_{\beta} =
-(-2\pi)^{-k-1}y^{-k-\frac{1}{2}}y^{\prime
-k-\frac{1}{2}}\Big((\sinh
\sigma)^{-1}\frac{d}{d\sigma}\Big)^{k}\cr\int_{\sigma}^{\infty}dr
\sinh(\frac{2\pi}{\beta}r) (\cosh r -\cosh
\sigma)^{-\frac{1}{2}}\pi(2\beta)^{-1}\Big(\cosh(\frac{2\pi}{\beta}r)
-\cos(\frac{2\pi}{\beta}(x_{0}-x_{0}^{\prime}))\Big)^{-1}
\end{array}\end{equation}
The Green functions for the massive Rindler model at finite
temperature (and for massive fields on the conical manifold) are
defined in eq.(27). From eqs.(39)-(40) it is easy to see that the
massless Green functions at $\beta=2\pi$ coincide with the ones of
the free quantum field on the Euclidean space as they should
because the metric (1) coincides with the flat metric in polar
coordinates when $x_{0}$ is periodic with the period $2\pi$. The
momentum of the finite temperature Rindler Green functions is
discrete  $p_{0}=2\pi l\beta^{-1}$ in eq.(36), where $l=0,\pm
1,...$ . Then, in eq.(35)
\begin{displaymath}
\nu=2\pi\vert l\vert \beta^{-1}
\end{displaymath}
For $\beta=4\pi\vert l\vert (2k+1)^{-1}$, where $k$ is a natural
number, the Legendre functions are expressed by elementary
functions of $z$ as in the case of the massless Green function on
the $D$ dimensional hyperbolic space.

\section{An outlook:analytic continuation and quantum fields}
             We can  discuss  now analytic continuations
 of the Green functions as functions of the Riemannian  metric (1).
The standard approach starts with a pseudoRiemannian metric
\cite{fulling}\cite{milne}. For a class of manifolds we can
construct quantum fields and calculate their Green functions.
These Green functions can be continued analytically to the
Riemannian metric (the Euclidean region). Starting from the
Riemannian metric we encounter some problems with an analytic
continuation, as discussed e.g. in \cite{wald}. There may be no
analytic continuation to a quantum field theory on a curved
background or the analytic continuation may be not unique. There
is no difficulty in the case of free fields defined on  static
space-times (the problem of an analytic continuation has been
solved in \cite{wald}). The interacting fields in the flat
space-time have an analytic continuation if their Euclidean
version is Osterwalder-Schrader (OS) positive ( then the Minkowski
version is Wightman positive). It has been noticed some years ago
(see \cite{dimock}\cite{jaffe} and references cited there) that
the reflection positivity of Green functions defined on Riemannian
manifolds allows to construct quantum fields on some
pseudoRiemannian manifolds although their physical meaning (in
particular the particle interpretation) may be obscure.

We discuss here solely analytic continuations of the Euclidean
Rindler model. First, we can continue, $ x^{0}\rightarrow ix^{0}$.
Then, we obtain the usual Rindler space. The continuation of Green
functions can be performed explicitly using eqs.(37)-(38) (and
(27) for the massive case). It can be seen directly from the Green
functions (37)-(38) that they are OS positive with respect to the
reflection $x_{0}\rightarrow -x_{0}$ (this is so because $(a^{2}
+(x_{0}-x_{0}^{\prime})^{2})^{-1}$ is OS posiitve). The analytic
continuation could also be performed by means of the operator
formalism as in ref.\cite{wald}(because the metric in the
Laplace-Beltrami operator (6) is time-independent). Next, we can
continue analytically $x_{1}\rightarrow ix_{1}$ (or equivalently
any $x_{k}$  with $k>1$).  In such a case we obtain a manifold
which is conformal to $R\times AdS$ (or to $S^{1}\times AdS$ for
the periodic Green functions (39)-(40)). Its Green function can be
obtained explicitly in the even case (37). The odd case (38) is
more complicated for negative (time-like) $z-1=\cosh\sigma -1$. In
such a case an analytic continuation of eq.(38) is needed. We may
use a definition of the Legendre function in terms of the
hypergeometric function for this purpose ( see \cite{grad}) or
consider the analytic continuation of the Legendre functions
directly from the integral representation \cite{kruczenski}. The
reflection positivity of Euclidean Green functions (37)-(38) and
the Wightman positivity of quantum fields (as its consequence) is
not obvious but shown in detail in \cite{jaffe}\cite{bros}.

 The third possibility is to continue
analytically $x_{0}\rightarrow ix_{0}$ simultaneously with
$y\rightarrow it$. In such a case we obtain the Milne space
\cite{milne}\cite{sommer}. An analytic continuation of eq.(37)
gives the Green function for the Milne space in the odd case. The
formula (38) for the even case needs  an analytic continuation (
$\sigma$ can be imaginary ) by means of the hypergeometric
function. Analytic continuation of eqs.(39)-(40) gives the formula
for the Green functions of the compactified Milne space discussed
in ref.\cite{turok}\cite{strom}\cite{russo}. However, it is not
clear whether the analytically continued Green functions satisfy
the Wightman positivity (i.e.,if the model defines quantum
fields).

There is another way to construct an analytic continuation using
the expansion in a complete set of solutions of the Klein-Gordon
equation. Then, the Euclidean Rindler Green function is expressed
in the form \cite{fulling}\cite{sessa}\cite{sommer}\cite{linet2}
\begin{equation}\begin{array}{l}
{\cal G}^{0}_{R}(x_{0},y,{\bf x};x_{0}^{\prime},y^{\prime},{\bf
x}^{\prime} )\cr =8 (2\pi)^{-D-1}\int dp_{1}d{\bf p}\sinh
(\pi\vert p_{1}\vert )\cr \exp(-\vert p_{1}\vert\vert
x_{0}-x_{0}^{\prime}\vert-i{\bf p}({\bf x}-{\bf
x}^{\prime}))K_{ip_{1}}(\vert{\bf
p}\vert\sqrt{y^{2}})\overline{K_{ip_{1}}(\vert{\bf
p}\vert\sqrt{y^{\prime 2}})}
\end{array}
\end{equation}where $K_{\nu}(y)=K_{-\nu}(y)$ is the modified  Bessel function of the third kind
of order $\nu$  \cite{grad} vanishing at infinity ; the square
root and the complex conjugation indicates the path of an analytic
continuation from $y>0$ to $y$ on the complex plane. The Green
function (41) is
 reflection positive with respect to the reflection $\theta
x_{0}=-x_{0}$, $\theta y=-y$ if we extend it from the Rindler
wedge to $y\leq 0$ by eq.(41). It is Wightman positive if the
analytic continuation $y\rightarrow it$ in eq.(41) is defined by
the formula
\begin{equation}
\exp(-ip_{1}( x_{0}-x_{0}^{\prime})-i{\bf p}({\bf x}-{\bf
x}^{\prime}))K_{ip_{1}}(i\vert{\bf p}\vert
t)\overline{K_{ip_{1}}(i\vert{\bf p}\vert t^{\prime })}
\end{equation}
The form of the two-point function resulting from the  analytic
continuation (42) coincides with the one which we  obtain if we
expand the Rindler quantum field in creation and annihilation
operators (see e.g. \cite{fulling}) and subsequently continue
analytically the modes. We may have difficulties with physical
interpretation of  quantum fields in a time-dependent metric. We
note however that in the Milne case at least in the limit
$t=\exp(u)\rightarrow 0$ ($u\rightarrow-\infty$ in eqs.(7)-(8))
the quantum field splits into the positive and negative frequency
parts ( as $K_{ip_{1}}(i\vert{\bf p}\vert t)\simeq \exp(ip_{1}u)$
if $u\rightarrow -\infty$; for some other quantum fields in a
time-dependent metric see \cite{ford}\cite{parker}). However, we
do not know whether the procedure of the analytic continuation
suggested here for the Milne model is unique. The resulting
two-point function may be different from the one which would have
come from an analytic continuation of eqs.(37)-(38). This problem
is still under investigation \cite{habaprep}.

 As a next step an interaction could be defined which determines
the S-matrix in terms of the propagators. The relation between
Rindler and (anti) de Sitter propagators (they are equal in the
momentum space; eq.(20)) indicates that the two approximations for
a near horizon geometry may lead to equivalent physical results
concerning scattering processes (however we should  keep in mind
that when calculating with hyperbolic propagators we must still
integrate over the mass).

{\bf Acknowledgements}:The author would like to thank an anonymous
referee for pointing out the relevance of the Killing bifurcate
horizons for the Rindler-type approximations.


\begin{thebibliography}{99}\bibitem{fulling}S.A. Fulling,
Phys.Rev.{\bf D7},2850(1973),
\newline
S.A. Fulling, J.Phys.{\bf A10},917(1977)

 \bibitem{milne} N.D. Birrell and P.C.W. Davies, Quantum Fields in
Curved Space, Cambridge University Press,1982
\bibitem{wald}R.M. Wald, Commun.Math.Phys.{\bf 70},221(1979)
\bibitem{dimock}J. Dimock, Rev.Math.Phys.{\bf 16},243(2004)
\bibitem{jaffe}A.Jaffe and G. Ritter,
Commun.Math.Phys,

{\bf 270},545(2007); arXiv:hep-th/0609003,



  \bibitem{bunch}T.S. Bunch and P.C.W. Davis,

   Proc.Roy.Soc.{\bf
  A360},117(1978)
  \bibitem{rindler}W. Rindler, Am. J.Phys.{\bf 34},1174(1966)

  \bibitem{ekpy}J. Khoury, B.A. Ovrut, N. Seiberg and N. Turok,


  Phys.Rev.{\bf D65},086007(2002)
\bibitem{horowitz}G.T. Horowitz and A. Strominger,

 Nucl.Phys.{\bf
B360},197(1991)\bibitem{maldacena}O. Aharony, S.S. Gubser, H.
Ooguri and Y. Oz,

Phys.Rep.{\bf 323},183(2000)


\bibitem{boulware}D.G.
Boulware,Phys.Rev.{\bf D11},1404(1975)
\bibitem{vandam}W. Troost and H.Van Dam, Nucl.Phys.{\bf
B152},442(1979)

\bibitem{moretti1}V.Moretti and L. Vanzo,

Phys.Lett.{\bf B375},54(1996); arXiv:hep-th/9507139



\bibitem{dowker1}J.S. Dowker, J.Phys.{\bf A10},115(1977)

\bibitem{dowker3}J.S. Dowker, Phys.Rev.{\bf D18},1856(1978)
\bibitem{dowker2} G. Cognola, K.Kirsten and L. Vanzo,

Phys.Rev.{\bf D49},1029(1994)

\bibitem{vilenkin}M. Barriola and A.Vilenkin, Phys.Rev.Lett.{\bf
63},341(1989)\bibitem{sereb}V.P. Frolov and E.M. Serebriany,


Phys.Rev.{\bf D35},3779(1987)

\bibitem{japan}K. Shiraishi and S. Hirenzaki,

 Class.
Quant.Grav.{\bf 9},2277(1992)
\bibitem{linet1}B. Linet,Phys.Rev.{\bf D35},536(1987)
\bibitem{linet2}B. Linet, arXiv:gr-qc/9505033

\bibitem{moreira}E.S. Moreira, Jr., Nucl.Phys.{\bf B451},365(1995)






\bibitem{fronsdal}C. Fronsdal, Phys.Rev.{\bf D10},589(1974)
\bibitem{ford}L.H. Ford and L. Parker,
Phys.Rev.{\bf D16},245(1977)
\bibitem{schom}Ch. Schomblond and P. Spindel, Ann.Inst.H.Poincare
{\bf 25},67(1976)\bibitem{mottola}E. Mottola, Phys.Rev.{\bf
D31},754(1984)
\bibitem{allen}B.Allen, Phys.Rev.{\bf D32},3136(1985)
 \bibitem{allenfolacci}
B. Allen and A. Folacci, Phys.Rev.{\bf D35},3771(1987)
\bibitem{kirsten}K. Kirsten and J. Garriga, Phys.Rev.{\bf
D48},567(1993)
\bibitem{tsamis} N.C. Tsamis
and R.P. Woodard,


Commun.Math.Phys.{\bf 162},217(1994)\bibitem{bros} J. Bros, H.
Epstein and U. Moschella,


Commun.Math.Phys.{\bf 231},481(2002)

\bibitem{rycher} N. Lohoue and  T. Rychener,

Commun.Math. Helv.{\bf 57},445(1982)

\bibitem{gaveau} A. Debiard and B.Gaveau, Can.J.Math.{\bf
39},1281(1987)
\bibitem{anker}J.-P. Anker and L. Li, Geom.func.anal.{\bf
9},1035(1999)

\bibitem{bif}
R.M. Wald, Quantum Field Theory in Curved Spacetime and Black Hole
Thermodynamics,University of Chicago Press,1994
 \bibitem{candelas} P. Candelas and J.S. Dowker,
Phys.Rev.{\bf D19},2902(1979)

\bibitem{kerner} A.A. Bytsenko, M.E.X. Guimaraes and R.
Kerner,

Eur.Phys.J.{\bf C39},519(2005);arXiv:hep-th/0501008
\bibitem{russo} J.G. Russo,Mod.Phys.Lett.{\bf A19},421(2004);

 arXiv:hep-th/0305032

A. Kehagias and J.G.Russo, JHEP {\bf 0007},027(2000);
arXiv/hep-th/0003281\bibitem{campo}R. Camporesi, Phys.Rep.{\bf
196},1(1990)
\bibitem{haba}Z.Haba,Phys.Rev.{\bf D38},
647(1988)

\bibitem{imperial}A.Grigoryan and M. Noguchi,


Bull.Lond.Math.Soc.{\bf 30},643(1998)
\bibitem{mckean}H.P. McKean, Comm.Pure
Appl.Math.{\bf 25},225(1972)
\bibitem{grad}I.S. Gradshtein and I.M.
Ryzhik, Table of Integrals, Series and Products, Academic
Press,New York,1962
\bibitem{bunch2}T.S. Bunch, Phys.Rev.{\bf D18},1844(1978)

\bibitem{temp}
S.A. Fulling and S.N.M.Ruijsenaars,

 Phys.Rep.{\bf
152},135(1987)
\bibitem{moretti2}V.Moretti,Class.Quant.Grav.{\bf 13},985(1996);arXiv/hep-th/9506142




\bibitem{kruczenski}U.H. Danielsson, E.
Keski-Vakkuri and M. Kruczenski,

JHEP,9901(1999)002;
 arXiv:hep-th/9812007






\bibitem{sommer}C.M. Sommerfield, Ann.Phys.{\bf 84},285(1974)
\bibitem{sessa}A. di Sessa, Journ.Math.Phys.{\bf 15},1892(1974)
\bibitem{turok}
  A.J. Tolley and N. Turok,

  Phys.Rev.{\bf D66},106005(2002);arXiv;hep-th/0204091

\bibitem{strom}G. T. Horowitz and A.R. Steif, Phys.Lett.{\bf
B258},91(1991) \bibitem{parker}

Ch. Charach and L. Parker, Phys.Rev.{\bf D24},3023(1981)
\bibitem{habaprep} Z. Haba, in preparation



\end{thebibliography}
\end{document}